\documentclass[10pt]{iopart}

\newcommand*{\papertitle}{Exploring room temperature spin transport under band gap opening in bilayer graphene}
\newcommand*{\affman}{Department of Physics and Astronomy, University of Manchester, M13 9PL, Manchester, UK}
\newcommand*{\affngi}{National Graphene Institute, University of Manchester, M13 9PL, Manchester, UK}
\newcommand*{\affecm}{Facultad de Ciencias Naturales y Matemáticas, Escuela Superior Politécnica del Litoral, ESPOL, Campus Gustavo Galindo, Km. 30.5 Vía Perimetral, P.O. Box 09-01-5863, 090902 Guayaquil, Ecuador}
\newcommand*{\affecd}{Center of Nanotechnology Research and Development (CIDNA), Escuela Superior Politécnica del Litoral, ESPOL, Campus Gustavo Galindo Km 30.5 Vía Perimetral, Guayaquil, Ecuador}
\newcommand*{\affconacyt}{Consejo Nacional de Ciencia y Tecnología (CONACyT), México}

\usepackage{graphicx}
\usepackage{bm}
\usepackage[breaklinks=true]{hyperref} 
\hypersetup{bookmarksnumbered, pdfpagemode=UseOutlines, pdfauthor={I.\ J.\ Vera-Marun}, pdftitle={\papertitle}, 
pdfdisplaydoctitle, colorlinks=true, citecolor=blue, filecolor=blue, linkcolor=blue, urlcolor=blue}
\usepackage{verbatim}
\usepackage{textcomp}
\usepackage{textgreek}             
\usepackage{iopams} 
\usepackage[justification=centering]{caption}

\begin{document}

\title{\papertitle}

\author{Christopher\ R.\ Anderson}
\address{\affman}
\ead{chris.anderson@manchester.ac.uk}
\vspace{10pt}

\author{Noel Natera-Cordero}
\address{\affman}
\address{\affconacyt}
\ead{noel.nateracordero@manchester.ac.uk}
\vspace{10pt}

\author{Victor H. Guarochico-Moreira}
\address{\affman}
\address{\affecm}
\address{\affecd}
\ead{vhuguaro@espol.edu.ec}
\vspace{10pt}

\author{Irina V. Grigorieva}
\address{\affman}
\address{\affngi}
\ead{Irina.V.Grigorieva@manchester.ac.uk}
\vspace{10pt}

\author{Ivan J. Vera-Marun}
\address{\affman}
\address{\affngi}
\ead{ivan.veramarun@manchester.ac.uk}
\vspace{10pt}

\begin{indented}
\item[]October 2022
\end{indented}

\begin{abstract}
We study the room-temperature electrical control of charge and spin transport in high-quality bilayer graphene, fully encapsulated with hBN and contacted via 1D spin injectors. We show that spin transport in this device architecture is measurable at room temperature and its spin transport parameters can be modulated by opening of a band gap via a perpendicular displacement field. The modulation of the spin current is dominated by the control of the spin relaxation time with displacement field, demonstrating the basic operation of a spin-based field-effect transistor. 
\end{abstract}

\maketitle

\section{Concept and novelty}

\begin{figure}[tbp]
    \centering
	\includegraphics*[angle=0, trim=0mm 0mm 0mm 0mm, width=83mm]{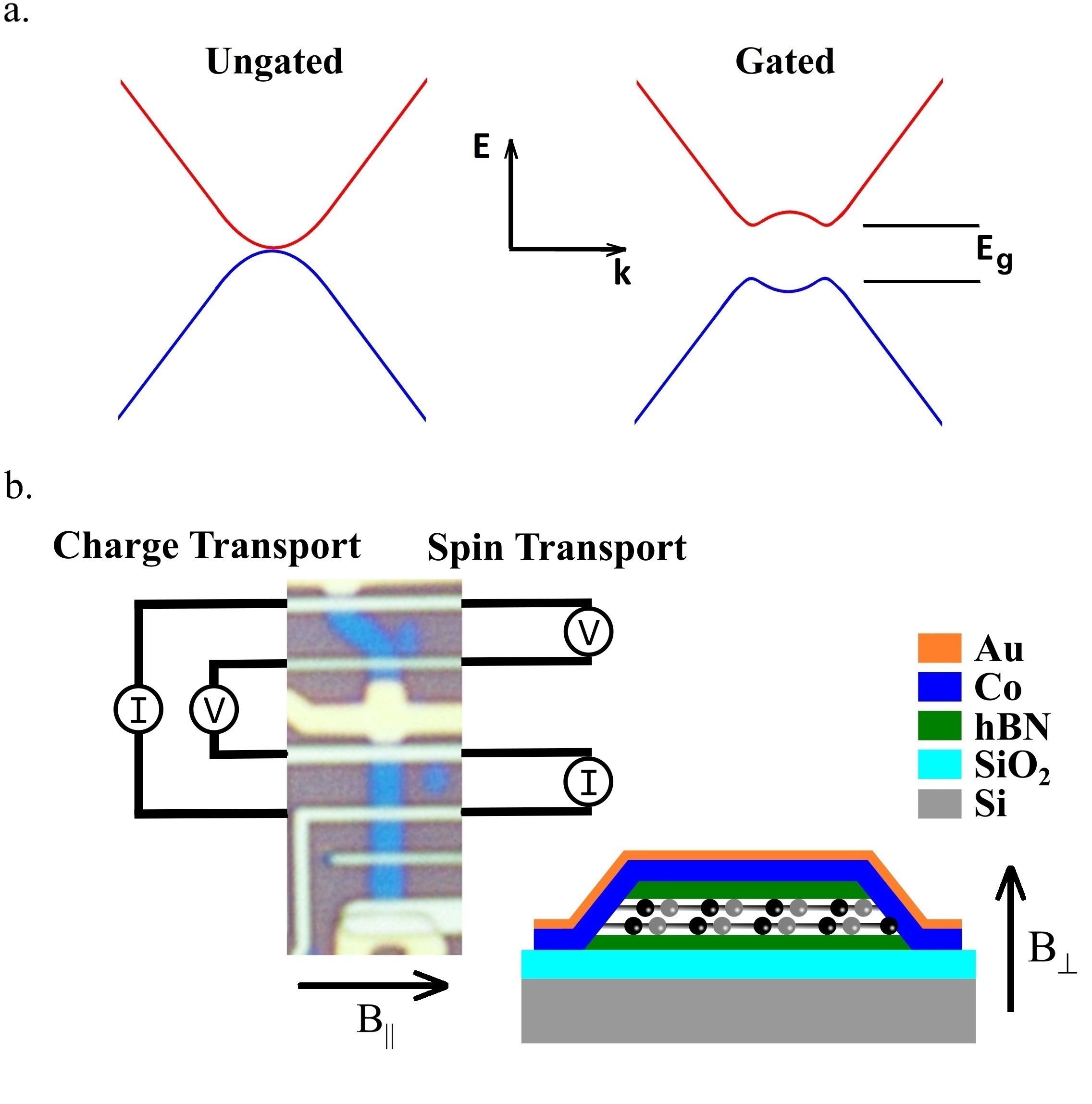}
	\caption{
		\label{fig:concept}
		\textbf{Bilayer graphene transport channel and device} \textbf{a.} Band structure of pristine BLG without (left) and with (right) an applied perpendicular electric displacement field \textbf{b.} Optical micrograph of of our $\sim$1\,\textmu m wide BLG graphene transport channels (blue) with contacts and top gate (in the measurement region). The charge (spin) transport current injection, $I$, and local (non-local) potential difference, $V$, measurement configuration. Inset: Schematic of the device heterostructure showing the Co/Au 1D edge contacts, with the graphene represented by the balls and sticks.}
\end{figure}

The state variable in digital devices has been realised, to date, by measurement of charge. Spintronics offers a new paradigm, whereby the quantum spin states of one or more electrons could be used as an alternative; improving performance by quick and efficient manipulation of that spin state \cite{flatte_semiconductor_2007}. Improvements in transport and control of spin are necessary for spin logic to be fully realised. Graphene offers desirable properties including theoretically large, although experimentally small spin `quality' parameters \cite{han_graphene_2014, RevModPhys.92.021003}. 
Bilayer graphene (BLG) presents the opportunity to electrically control its band gap \cite{Zhang2009, Taychatanapat2010}, $E_g$, (see Figure \ref{fig:concept}.a.) and, as a result, the spin transport parameters in the channel. Graphene based spin-based field effect transistors have been previously attempted: Materials with large and electrostatically tunable spin-orbit coupling (SOC) have been brought into contact with a graphene channel, enabling spins to be absorbed \cite{Yan2016b, Dankert2017}; this precludes the fabrication of a fully encapsulated homogeneous transport channel. A combination of injector current and back gate voltage can be tuned to create a spin transistor-like action by reversing the polarity of the contact spin polarisation \cite{Ringer2018, Xu2018}, however, the latter varied between contacts and consequently so did the spin signal, making this approach not scalable. High-quality graphene devices with reproducible contact spin polarisation can overcome this limitation \cite{Guarochico-Moreira2022}. 
Alternatively, the electrostatic control of the graphene channel's band gap \cite{Avsar2016, Ingla-Aynes2015, Ingla-Aynes2021} offers a small, but promising effect, which has the potential to be scalable whilst maintaining a simple architecture, with excellent transport properties.

Spin transport has previously been observed in single layer (SLG) and BLG devices which have transparent (invasive) contacts (exhibiting the spin resistance mismatch problem \cite{Takahashi2003, Maassen2012}) and a SiO\textsubscript{2} substrate \cite{Xu2018b, Tombros2007, Han2010} which brakes graphene's inversion symmetry causing Rashba fields, increased SOC and spin relaxation. Alternatively, devices have been fabricated with an hBN substrate but are not fully encapsulated \cite{Ingla-Aynes2015}, leading to a heterogeneous channel consisting of sections with and without encapsulation that suffer from environmental degradation. Quantum tunnelling contacts have been shown to solve the resistance mismatch problem \cite{Tombros2007, Avsar2016, Ingla-Aynes2015, Gurram2016a, Leutenantsmeyer2018} improving the measured spin transport parameters. However, such tunneling contacts are difficult to fabricate due to troublesome selection of atomically thin hBN \cite{Gorbachev2011} or inconsistency of the growth of an atomically thin barrier \cite{Tombros2007}.

The effect which we have observed, where we propose that spin anisotropy is induced in the BLG by a small but tuneable band gap, has previously been seen but at high carrier densities and low temperatures \cite{Xu2018b}. In contrast, here we experimentally study a bilayer device at room temperature (RT) and show how,  close to the Dirac point, the transport parameters are effected by an applied perpendicular displacement field which induces the band gap to open (Figure \ref{fig:concept}.a). In this regime, however, it is typically difficult to measure the device's spin transport parameters, as a consequence of the low signal to noise ratio and an induced band gap commensurate with the thermal activation energy.
Conversely, at the Dirac point the effects of the band gap opening on spin transport are more readily observed, with the spin splitting being more prominent at the band edge. Consequently, the quality of the homogeneous channel is of paramount importance, which allows the observation of a modulation of the spin relaxation time, $\tau_s$, which has not previously been shown close to the Dirac point and at RT. Here this is made possible by the fabrication of a BLG transport channel which is fully encapsulated, where the effect of the contacts and the resistance mismatch problem are minimised by the use of nanoscale 1D edge contacts \cite{Xu2018, Guarochico-Moreira2022}. 

\section{Results}

\subsection{Device architecture and fabrication}
The full encapsulation of the BLG with hBN leaves only the sandwiched graphene's edges exposed; on these edges 1D contacts are formed (Figure \ref{fig:concept}.b). The device was designed to enable electrostatic gating of the whole device using the Si substrate as a back gate and of a region between two contacts using a microfabricated top gate. The design facilitates the fabrication of multiple fully encapsulated regions with top gate electrodes. The 1D edge magnetic contacts are fabricated from gold-capped cobalt, whilst the electrostatic top gates are gold, with a thin chromium adhesion layer.

\subsection{Charge transport measurements}
\begin{figure}[tbp]
    \centering
	\includegraphics*[angle=0, trim=12mm 0mm 0mm 0mm, width=85mm]{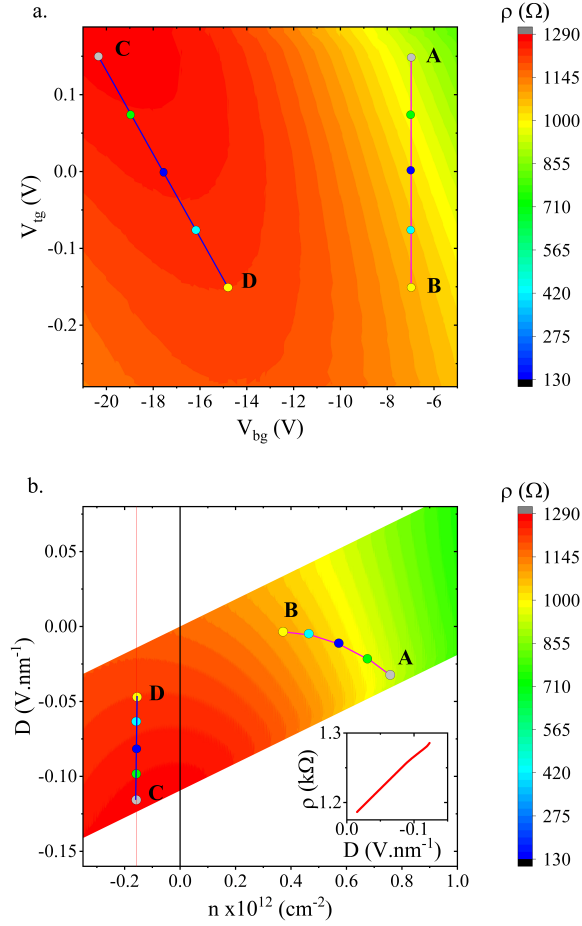}
	\caption{\textbf{a.} A 2D charge transport measurement map at RT showing the effect on the sheet resistance, $\rho$, of applying a back, $V_{bg}$, and top gate voltage, $V_{tg}$. \textbf{b.} The same charge transport  measurements transformed into a map as a function of carrier density, $n$, and electric displacement field, $D$. In both maps the symbols are shown where the spin transport measurements were made. \textbf{b. (inset)} Sheet resistance close and parallel to the Dirac ridge ($n \approx 0$)}
    \label{2D Dirac map + SV measurement points}
\end{figure}

Our local measurements of the top-gated and hBN-encapsulated BLG, using the configuration shown in Figure \ref{fig:concept}.b, explored the charge transport by varying the back gate, $V_{bg}$, and top gate, $V_{tg}$, voltages and measuring the BLG resistance, $R$. From the latter the sheet resistance, $\rho$, is calculated and exhibits a modulation as a function of the gate voltages as shown in Figure \ref{2D Dirac map + SV measurement points}.a. 
The gate voltages are used to calculate the carrier density, $n$, and electrical displacement field, $D$, which enable re-plotting the sheet resistance  in terms of $n$ and $D$. Figure \ref{2D Dirac map + SV measurement points}.a and b show this transformed map at RT, along with symbols indicating a selection of ($n, D$) pairs where spin transport measurements were performed. The inset of Figure \ref{2D Dirac map + SV measurement points}.b shows a near linear relationship between the transport channel's sheet resistance and the applied displacement field, close and parallel to the Dirac ridge (the path of zero carrier density as the displacement field is modulated), exhibiting a monotonic increase in $\rho$ as $D$ increases in magnitude.

The device shows a carrier field effect mobility, $\mu$, of up to 40,000 cm\textsuperscript{2}/Vs at 20 K and 11,000 cm\textsuperscript{2}/Vs at RT; where $\mu = (d\sigma/{dn})/e$ and $\sigma = 1/\rho$ and are evaluated at a moderate carrier density, $|n| \sim 1 \times 10^{12}$\,cm\textsuperscript{-2}. These values are comparable with those achieved in fully encapsulated BLG, tunnel barrier contacted devices \cite{Avsar2016, Gurram2016a, Leutenantsmeyer2018} and substantially higher than non-encapsulated BLG devices \cite{Ingla-Aynes2015, Xu2018b}. 
The BLG transport channel does not suffer from any significant contact doping due to the small contact area (a few carbon atoms depth from the edge of the device) and small contact width ($\le400$ nm) \cite{Guarochico-Moreira2022} and minimal bubbles within the heterostructure. We propose that even higher mobilities may be obtained with the fabrication of a wider channel and eradication of any residual bubbles in the channel, potentially improving the spin transport parameters of the device, however, this is dependent on the cause of the scattering \cite{Avsar2016}. We have confirmed that the top gate voltage has little effect on the resistance of the regions either side of the top gated region.

\subsection{Spin transport measurements}
\begin{figure}[tbp]
    \centering
	\includegraphics*[angle=0, trim=10mm 0mm 0mm 0mm, width=85mm]{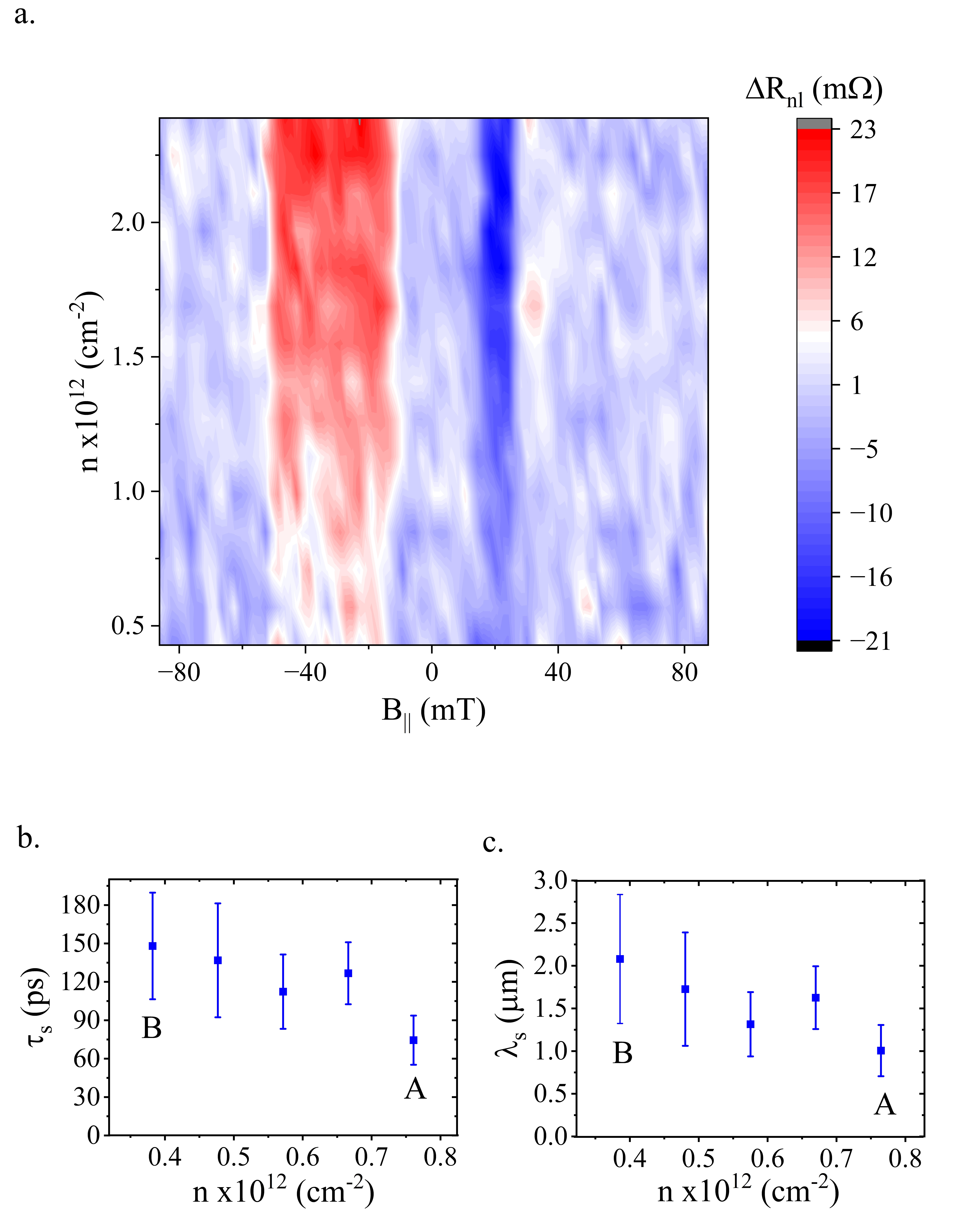}
	\caption{\textbf{a.} 2D map of spin valves at RT as the electron carrier density is increased. The anti-parallel magnetic alignments are clearly visible as the vertical red and darker blue bands. As the CNP is approached the signal becomes increasingly weak. \textbf{b.} and \textbf{c}. Spin transport parameters ($\tau_s$ and $\lambda_s$, respectively) extracted from spin precession measurements at RT, for the 5 points along the A-B path shown in Figure \ref{2D Dirac map + SV measurement points}.
	}
	\label{2d spin transport map & Tau_s and Lambda_s @-7Vbg}
\end{figure}

To characterise spin transport we first measured a series of spin valves, using the non-local configuration shown in Figure \ref{fig:concept}.b, at RT. These are measurements where we sweep an in-plane magnetic field, $B_\parallel$, applied along the direction of contacts, which reverses the magnetisation of the 1D contacts and enables us to obtain either a parallel or antiparallel magnetic alignment between the injector and detector contacts. The trace and retrace scans of magnetic field are then subtracted from one another to obtain a purely spin-dependent signal. Such measurements were repeated across a carrier density range ($n$ = 0.45 to 0.8 \texttimes 10 \textsuperscript{12}\,cm\textsuperscript{-2}), with the top gate grounded, the results of which are shown in Figure \ref{2d spin transport map & Tau_s and Lambda_s @-7Vbg}.a. The vertical bands of colour seen to the left (red) and right (darker blue) of zero magnetic field strength, $B = 0$ T, are the distinctive spin-valve signals as the magnetic injector/detector configurations change from parallel to anti-parallel. The average magnitude of the spin signal, the difference in non-local resistance between parallel and anti-parallel magnetic configurations, $\Delta R_{nl}$, across this range of $V_{bg}$ is $\sim$10 m\textOmega. 
A close examination of the 2D map shows that the spin signal and the signal to noise ratio (SNR) diminish as the back gate approaches the charge neutrality point (CNP, $n$ = 0) from the electron carrier regime. This is a characteristic of 1D edge contacts, whose spin polarisation approaches zero near the CNP \cite{Xu2018}. 

\begin{figure}[tbp]
    \centering
	\includegraphics*[angle=0, trim=12mm 0mm 0mm 0mm, width=85mm]{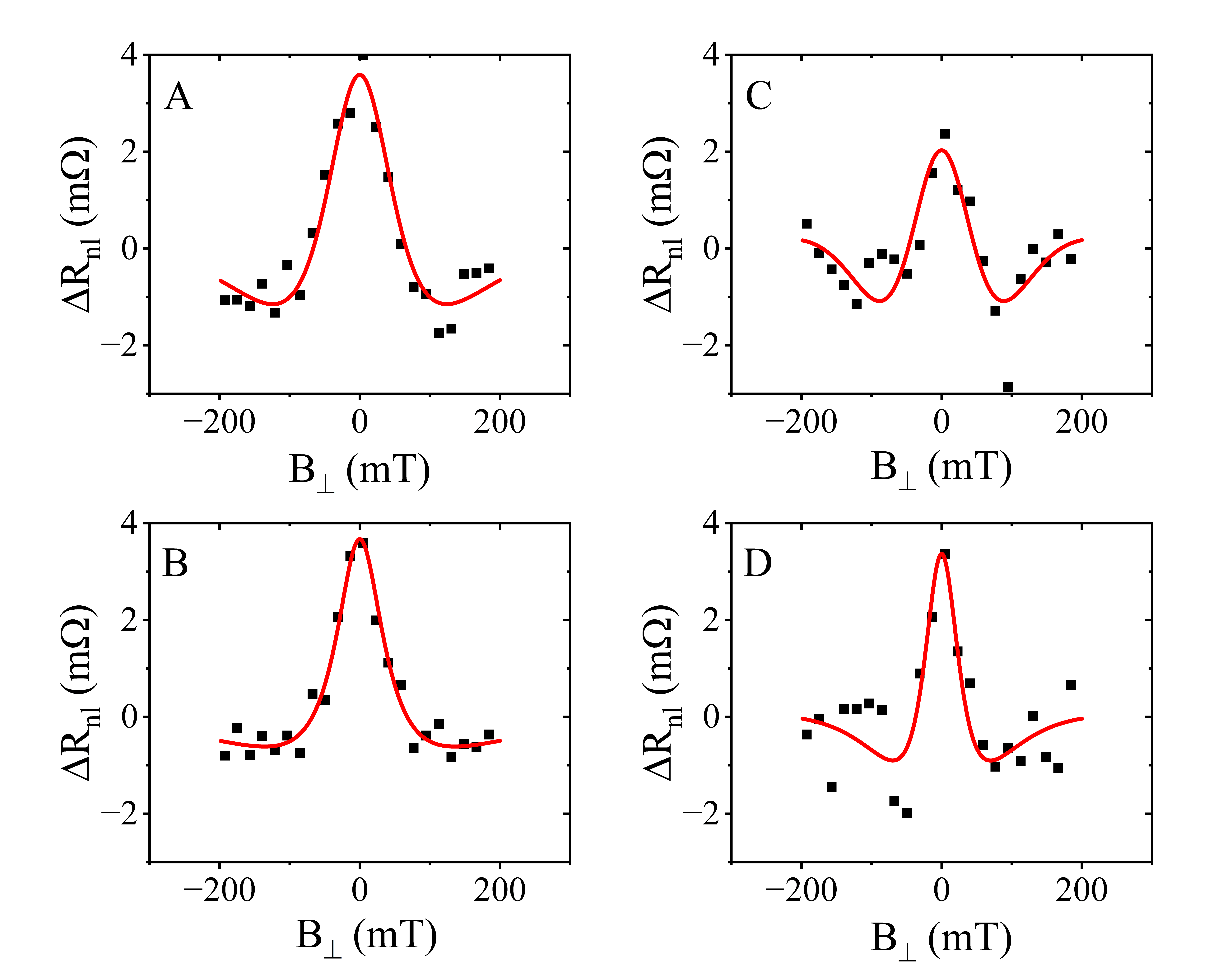}
	\caption{Spin precession measurements at RT. The lineshape of the Hanle curve changes with the change in $n$, shown in A and B, whilst C and D (at $n \approx$ 0) shows the lineshapes when the displacement field changes. The labels correspond to the same ($n, D$) pairs labelled in Figure \ref{2D Dirac map + SV measurement points}. The Hanle equation fits are shown in red.
	}
	\label{Hanle}
\end{figure}

The changes in spin signal as a function of carrier density were further studied by performing spin precession measurements. In doing so, we sweep a magnetic field strength perpendicular to the plane of the graphene (see Figure \ref{fig:concept}.b), $B_\bot$, across a range of $\pm$200 mT, which causes the diffusing electronic spins to experience Larmor precession, and fit the spin signal with the Hanle equation \cite{Johnson1988}. The spin relaxation time, the spin diffusion coefficient and the corresponding spin relaxation length, are extracted from the Hanle curve fits (examples shown in Figure \ref{Hanle}). Uncertainties in determining these parameters are realised by the error bars in Figures \ref{2d spin transport map & Tau_s and Lambda_s @-7Vbg}.b, c and \ref{Eg vs D, DRnl, Tau_s and Lambda_s @n=0}. We use the spin, rather than the charge diffusion coefficient in our analysis, as the measurements are made close to the neutrality point, whilst also opening a band gap, consequently it is inappropriate to use the Einstein relation to extract the charge diffusion coefficient from charge transport measurements. We select five pairs of ($n, D$) values for spin transport measurements in the electrons regime, at fixed top gate voltage and $-7$~V\textsubscript{bg}, where carrier density variation dominates over variation in displacement field. The selected five  points are shown in Figure \ref{2D Dirac map + SV measurement points} (A-B path). The corresponding spin transport measurements and Hanle fits are shown in \ref{Hanle} (A and B) and the extracted spin transport parameters are shown in Figure \ref{2d spin transport map & Tau_s and Lambda_s @-7Vbg}.b and c. Here we note that $\tau_s$ and ${\lambda_s}$ decrease with increasing carrier density (at low carrier density) which, although small, is commensurate with that typically observed for BLG spin devices \cite{Avsar2016}.

Next, we explore spin transport near charge neutrality under the effect of a displacement field. For this purpose we measured spin precession at five different ($n,D$) values indicated by the C-D path shown in Figure \ref{2D Dirac map + SV measurement points}.a and b. Here we fixed a carrier density close to the CNP, $n<2\times 10^{11}$\,cm\textsuperscript{-2}, and varied $D$ from $-0.04$\,V\,nm\textsuperscript{-1} to $-0.12$\,V\,nm\textsuperscript{-1}. Examples of the spin precession measurements are shown in Figure \ref{Hanle} C and D. A wider peak is seen at the high end of the displacement field range (C) and a narrower peak at the low end of the range (D). These changes in lineshape correspond to a modulation in the spin transport parameters extracted from the Hanle analysis. The spin transport parameters: spin relaxation time, $\tau_s$, spin relaxation length, $\lambda_s$, and spin signal (at $B$ = 0), are shown in Figure \ref{Eg vs D, DRnl, Tau_s and Lambda_s @n=0}.a, b and c, respectively. As the magnitude of $D$ is increased there is a threefold decrease in $\tau_s$, and a corresponding modulation of $\lambda_s$. 

\begin{figure}[tbp]
    \centering
	\includegraphics*[angle=0, trim=0mm 0mm 0mm 0mm, width=85mm]{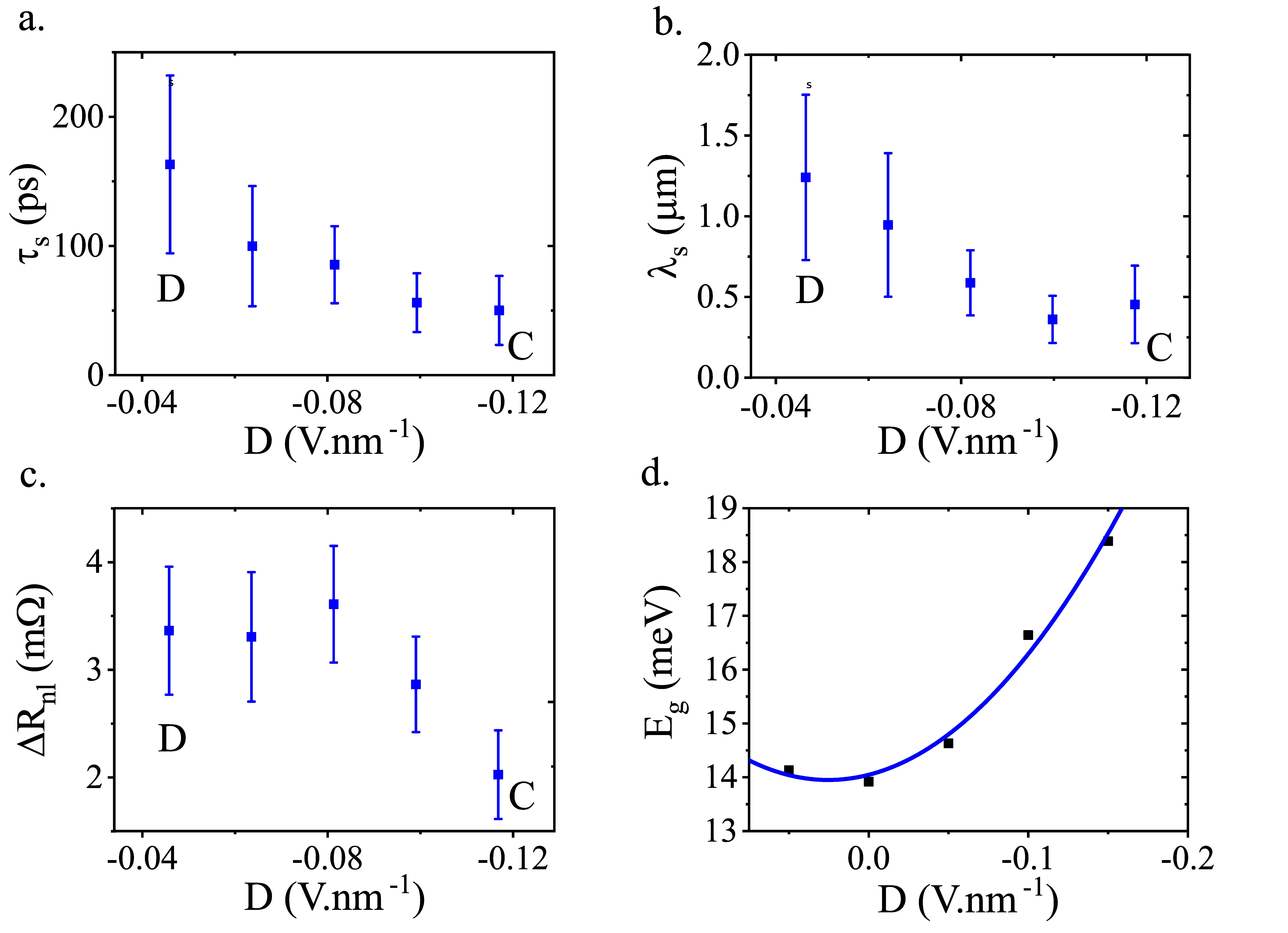}
	\caption{Spin transport measurement and analysis close and parallel to the Dirac ridge at RT. Spin relaxation time (\textbf{a}), spin relaxation length (\textbf{b}) and spin signal (\textbf{c}) for the five points along the C-D path shown in Figure \ref{2D Dirac map + SV measurement points}. \textbf{d.} Band gap energy, $E_g$, vs displacement field, $D$, from Arrhenius analysis of charge transport measurements.}
	\label{Eg vs D, DRnl, Tau_s and Lambda_s @n=0}
\end{figure}

\section{Discussion}

The spin signal, $\Delta R_{nl}$, measured along the Dirac ridge ($n\sim0$) at RT, shows a decrease with increasing displacement field (Figure \ref{Eg vs D, DRnl, Tau_s and Lambda_s @n=0}.c), with a modulation of approximately half.

To understand the role of the applied displacement field, $D$, we measured a series of maps of local resistance versus $n$ and $D$ from 20\,K up to RT. This enables the study of thermal excitation of carriers for a given $D$, with a fixed $n\approx0$. Arrhenius analysis \cite{Arrhenius1889a, Arrhenius1889, Xu2018b} of these measurements yielded the value of the band gap in the encapsulated BLG channel, exhibiting band gap values of up to $\sim 19$\,meV as the magnitude of the displacement field increases to 0.2\,V\,nm\textsuperscript{-1} (see Figure \ref{Eg vs D, DRnl, Tau_s and Lambda_s @n=0}.d).
A finite band gap at zero displacement field applied is noted, 
along with a non-linear relationship between the magnitude of $D$ and $E_g$, which may be caused by the presence of the hBN encapsulation. These results are commensurate with previous work on BLG \cite{Zhang2009, Xu2018b}.

Having evidenced the role of the displacement field in opening a band gap in the BLG channel, we proceed to consider how this relates to the changes observed in spin transport at RT. We note that the dependence of $\tau_S$, $\lambda_S$ and $\Delta R_{nl}$ on $D$ is correlated with the shape of the relationship between $E_g$ and $D$. At low $D$ there is little change in $E_g$; the larger changes in $E_g$ only being seen at higher $D$ (Figure \ref{Eg vs D, DRnl, Tau_s and Lambda_s @n=0}.d). Therefore, within the transport parameter error bars shown (Figure \ref{Eg vs D, DRnl, Tau_s and Lambda_s @n=0}.a-c), we propose that for \mbox{$|D| \lesssim$ 0.08\,V.nm\textsuperscript{-1}} reveal no effect of the displacement field, however, for \mbox{$|D|$ > 0.08\,V.nm\textsuperscript{-1}} we begin to see an effect. A displacement field, beyond the range shown, was limited by potential damage to the device.  

At RT and close to the Dirac point it is typically difficult to measure the device's spin transport parameters, as a consequence of the low signal to noise ratio, the resistance mismatch problem \cite{Fabian2007} and an induced band gap which is commensurate with the thermal activation energy. However, at the Dirac point the effects of the band gap are more readily observed, with the spin splitting being more prominent at the band edge in BLG \cite{Konschuh2012}.

It should be noted that our measurement configuration is designed to evaluate the difference in energy between the spin up and spin down directly, rather than attempting to measure the spin up and spin down accumulations independently. Consequently, it is sensitive to small net spin accumulations, of the order of $\sim$ \textmu eV. Also, the quality of our homogeneous channel allows us to show a modulation of the spin relaxation time, $\tau_s$, which has not previously been shown. This is made possible by the fabrication of a BLG transport channel which is fully encapsulated, where the effect of the contacts and the resistance mismatch problem are minimised by the use of nanoscale 1D edge contacts \cite{Xu2018, Guarochico-Moreira2022}.

A trivial mechanism by which opening a band gap can modulate spin transport is via the modulation of the charge resistance. An increase of the sheet resistance, $\rho$, in the region where the displacement field is applied, between the injector and detector contacts, would inhibit carrier diffusion and prevent the flow of spin current across this region \cite{Avsar2016}. We indeed observe a modulation of charge transport at RT, as shown in the inset of Figure \ref{2D Dirac map + SV measurement points}.b, confirming that the band gap is being electrically opened. Nevertheless, at RT this only accounts for a change of 5\% in $\rho$ for the spin transport measurements along the C-D path. 
The spin diffusion between injector and detector contacts, at a distance $L = 3.0$\,\textmu m apart, is described by the following relation \cite{Tombros2007},
\begin{equation} \label{DRnl}
    \Delta R_{nl} \propto \rho \lambda_S e^{-\frac{L}{\lambda_S}},
\end{equation}
from which we would expect, to first order, the spin signal, $\Delta R_{nl}$, to only increase by 5\% with increasing displacement field, as it is proportional to $\rho$.
On the contrary, we observe that $\Delta R_{nl}$ decreases as the displacement field increases (see Figure \ref{Eg vs D, DRnl, Tau_s and Lambda_s @n=0}.c), with an on/off ratio of $\lesssim 2$ from D to C.

A more significant mechanism by which the displacement field can modulate spin transport is via changes in the spin relaxation length, $\lambda_S$. As shown in Equation (\ref{DRnl}), the factors containing $\lambda_S$ have the effect of decreasing $\Delta R_{nl}$ as $\lambda_S$ decreases, which is consistent with the observed dependence of $\lambda_S$ on displacement field (see Figure \ref{Eg vs D, DRnl, Tau_s and Lambda_s @n=0}.b). Therefore the effect of $\lambda_S$ overwhelms the small effect of $\rho$. Note that ideally, based on a twofold decrease in $\lambda_S$, one would expect a $\Delta R_{nl}$ on/off ratio of up to 200, larger than that observed. We find that this expected modulation in $\Delta R_{nl}$ is, however, muted in our measurements, due to a simultaneous increase in the contact polarisation from 0.5\% to 5\%, which partially counteracts the effect of $\lambda_S$. The latter is a limitation of our current device design, where the contacts are under the influence of the back gate, and is consistent with a strong dependence of the polarisation of 1D contacts as a function carrier density \cite{Xu2018}. Modification of the device design, by introducing a local back gate that does not act on the contacts, would enable a larger on/off ratio due to only the effect of opening a band gap. Given that the charge transport parameters, resistivity and diffusion coefficient, vary only by 5\% across our range of displacement field, the variation in the spin relaxation length is given by the dependence \mbox{$\lambda_S \propto \sqrt{\tau_s}$}.

We observe that as the displacement field is increased, the spin relaxation time, $\tau_s$, decreases by up to a threefold, and correspondingly the length, $\lambda_s$ decreases too. 
This observation is commensurate with the increasing width and decreasing height of the central peak in the spin precession measurements (Figure \ref{Hanle}) as $D$ increases. 
Previous work has demonstrated that in BLG the spin lifetime anisotropy, the ratio of the out-of-plane to the in-plane spin relaxation time, is modulated by a perpendicular displacement field, due to the resulting out-of-plane spin-orbit fields \cite{Xu2018b}. This previous work showed that the anisotropy increases with increasing displacement field in BLG, as a result of both the increase in out-of-plane spin relaxation time and the decrease in the in-plane spin relaxation time. However, unlike our experiment, this was carried out at high carrier densities and low temperatures. 

Given our measurement geometry determines the transport of in-plane spins \cite{Tombros2007}, such a change in anisotropy due out-of-plane spin-orbit fields in BLG is commensurate with our observations shown in Figure \ref{Eg vs D, DRnl, Tau_s and Lambda_s @n=0}.a. 
Note that this result of a perpendicular displacement field is opposite to that present in SLG, where it results in an effective in-plane Rashba spi-orbit field that causes dephasing of the out-of-plane spins, i.e.\ the opposite effect on spin anisotropy \cite{Guimaraes_Controlling_2014}. On the other hand, in BLG the dominant phenomena is that of the change in band structure due to opening the band gap, which allows spin splitting at the band edges and gives rise to the observed  modulation of spin relaxation time with displacement field \cite{Xu2018b}. 
The spin transport measurements and analysis described above, presented in Figures \ref{Eg vs D, DRnl, Tau_s and Lambda_s @n=0}.b, c and d, show that it is possible to modulate the spin signal in a BLG spin transistor operating at RT, by virtue of the resulting modulation in $\tau_s$ when opening the band gap.

\section{Conclusion}
We have shown that in a fully hBN encapsulated, 1D edge contacted BLG at RT, it is possible to electrostatically modulate a nonlocal spin current purely by a displacement field. The latter contributes to opening a band gap, with only a small modulation of the charge transport, but a sizeable modulation of the spin relaxation time and corresponding spin signal. 
 
This work, using a van der Waals device architecture which ensures minimally invasive contacts and scalable design of electrostatic gates, paves the way towards further exploration of bilayer graphene as a prototypical spin-FET at RT, which has application in spin logic \cite{Sierra2021}. SOC effects are now of significant interest in technological areas of spintronics such as for memory devices \cite{Yang2022c}. Further, investigation of the combination of proximity induced SOC with the tuneable band gap SOC afforded by BLG and/or a drift current \cite{Ingla-Aynes2018a} may produce an enhanced on/off ratio and spin current guiding.

\section{Methods}

\subsection{Sample Fabrication}

BLG samples were prepared by mechanical exfoliation of natural graphite \cite{Novoselov2004}, using the so called Scotch tape method and transferred onto a Si/SiO\textsubscript{2} (290 nm) substrate. BLG candidates were initially found using optical microscopy and the number of layers confirmed with Raman spectroscopy. Similarly hBN candidates were identified, confirming their thickness with a combination of AFM and profilometry. The bottom hBN flake of our devices has a thickness of 12.5 nm and the top flake 5.5 nm. Using the dry transfer technique, a Poly Methyl Methacrilate (PMMA) membrane ($\sim$500 nm thick) lifted the top hBN flake, assisted by a transfer rig equipped with micromanipulators and a hot plate, which improved the membrane stickiness. This top hBN, on the membrane, was used to lift the BLG with van der Waals forces. Finally both flakes were deposited on top of the bottom hBN, already on the Si/SiO\textsubscript{2} substrate.  The remaining PMMA membrane was removed by soaking the sample in acetone and IPA for $\sim$5 minutes each. The stack was annealed in a H/Ar atmosphere for 3 hours at 300 \textdegree C to remove PMMA residues and contamination. AFM imaging qualitatively determined the cleanest areas of the stack (clean, bubble free) where the graphene spin channel could be patterned. A double PMMA electron beam lithography (EBL) resist layer (solutions of 3\% of each of 495 and 950 molecular weight) was spin-coated onto the sample. The channel was patterned with EBL and etched with an O\textsubscript{2}/CHF\textsubscript{3} plasma, leaving the graphene's edges exposed, ready to accept 1D edge contacts. Finally, contacts were patterned using EBL and Cr/Au evaporated to form non-magnetic contacts, whilst Co/Au was used to form ferromagnetic contacts with a range of aspect ratios. 

\subsection{Transport Measurement}
All spin transport measurements were made in a standard non-local lateral 4 point configuration. Spin precession measurements were fitted with the Hanle equation and spin parameters extracted, the results of which are shown in Figures \ref{Eg vs D, DRnl, Tau_s and Lambda_s @n=0}.b, c and d. The measurements were made by setting the contact magnetic configuration to parallel (P) and anti-parallel using an in-plane magnetic field in a non-local spin valve configuration. In each of the parallel and anti-parallel configurations an out of plane magnetic field was swept and the non-local resistance, $R_{nl}$, recorded. The top and back gate voltages, $V_{tg}$ and $V_{bg}$, were stepped to achieve the required carrier density, $n$, and electric displacement field, $D$, according to Equations (\ref{top and back gate n}) and (\ref{top and back gate D}) \cite{Xu2018b},

\begin{equation} \label{top and back gate n}
    n = \epsilon_0 \epsilon_{r_{bg}} \frac{V_{bg} - V_{D_{bg}}}{e t_{bg}} + \epsilon_0 \epsilon_{r_{tg}} \frac{V_{tg} - V_{D_{tg}}}{e t_{tg}}
\end{equation}

\begin{equation} \label{top and back gate D}
    D = \epsilon_{r_{bg}} \frac{V_{bg} - V_{D_{bg}}}{2 t_{bg}} + \epsilon_{r_{tg}} \frac{V_{tg} - V_{D_{tg}}}{2 t_{tg}}
\end{equation}
where $V_{D}$ is the top gate, $tg$, and back gate, $bg$, voltage required to overcome residual channel doping and achieve charge neutrality. $t$ is the thickness of the gate dielectric and $\epsilon_{r}$ is the relative permittivity of the gate dielectric, which can be calculated from the respective gate's capacitance, $C$, equations,
\begin{equation} \label{back gate capacitance}
    C_{bg} = \frac{\epsilon_{r_{bg}}}{t_{bg}} A = \frac{\epsilon_{r_{SiO_2}} \epsilon_{r_{bg hBN}}}{\epsilon_{r_{SiO_2}} t_{bg hBN} + \epsilon_{r_{gb hBN}} t_{SiO_2}} A
\end{equation}
\begin{equation} \label{top gate capacitance}
    C_{tg} = \frac{\epsilon_{r_{tg}}}{t_{tg}} A =  \frac{\epsilon_{r_{tg hBN}}}{t_{tg hBN}} A
\end{equation}
where $A$ is the area of the gate, ${\epsilon_{r_{SiO_2}}}$ = 3.9 and ${\epsilon_{r_{bg hBN}}}$ = ${\epsilon_{r_{tg hBN}}}$ = 3.2 are the relative permittivities of the SiO\textsubscript{2} substrate and hBN, respectively. $t_{tg hBN}$ and $t_{bg hBN}$ are the thicknesses of the two hBN flakes, described earlier. The device presented here has a SiO\textsubscript{2} thickness, ${t_{SiO_2}}$ = 290 nm.

\section*{Data availability}
The datasets used and/or analysed during the current study are available from the corresponding author on reasonable request.

\section*{References}
\bibliographystyle{_naturemag-Ivan}
\bibliography{references}

\begin{thebibliography}{10}
\expandafter\ifx\csname url\endcsname\relax
  \def\url#1{\texttt{#1}}\fi
\expandafter\ifx\csname urlprefix\endcsname\relax\def\urlprefix{URL }\fi
\providecommand{\bibinfo}[2]{#2}
\providecommand{\eprint}[2][]{\url{#2}}

\bibitem{flatte_semiconductor_2007}
\bibinfo{author}{Flatt{\'e}, M.~E.}
\newblock \bibinfo{title}{Semiconductor spintronics for quantum computation}.
\newblock In \bibinfo{editor}{Flatt{\'e}, M.~E.} \&
  \bibinfo{editor}{{\c{T}}ifrea, I.} (eds.)
  \emph{\bibinfo{booktitle}{Manipulating Quantum Coherence in Solid State
  Systems}}, \bibinfo{pages}{1--52} (\bibinfo{publisher}{Springer Netherlands},
  \bibinfo{address}{Dordrecht}, \bibinfo{year}{2007}).

\bibitem{han_graphene_2014}
\bibinfo{author}{Han, W.}, \bibinfo{author}{Kawakami, R.~K.},
  \bibinfo{author}{Gmitra, M.} \& \bibinfo{author}{Fabian, J.}
\newblock \bibinfo{title}{Graphene spintronics}.
\newblock \emph{\bibinfo{journal}{Nat Nano}} \textbf{\bibinfo{volume}{9}},
  \bibinfo{pages}{794--807} (\bibinfo{year}{2014}).

\bibitem{RevModPhys.92.021003}
\bibinfo{author}{Avsar, A.} \emph{et~al.}
\newblock \bibinfo{title}{Colloquium: Spintronics in graphene and other
  two-dimensional materials}.
\newblock \emph{\bibinfo{journal}{Rev. Mod. Phys.}}
  \textbf{\bibinfo{volume}{92}}, \bibinfo{pages}{021003}
  (\bibinfo{year}{2020}).

\bibitem{Zhang2009}
\bibinfo{author}{Zhang, Y.} \emph{et~al.}
\newblock \bibinfo{title}{{Direct observation of a widely tunable bandgap in
  bilayer graphene}}.
\newblock \emph{\bibinfo{journal}{Nature}} \textbf{\bibinfo{volume}{459}},
  \bibinfo{pages}{820--823} (\bibinfo{year}{2009}).

\bibitem{Taychatanapat2010}
\bibinfo{author}{Taychatanapat, T.} \& \bibinfo{author}{Jarillo-Herrero, P.}
\newblock \bibinfo{title}{Electronic transport in dual-gated bilayer graphene
  at large displacement fields}.
\newblock \emph{\bibinfo{journal}{Phys. Rev. Lett.}}
  \textbf{\bibinfo{volume}{105}}, \bibinfo{pages}{166601}
  (\bibinfo{year}{2010}).

\bibitem{Yan2016b}
\bibinfo{author}{Yan, W.} \emph{et~al.}
\newblock \bibinfo{title}{{A two-dimensional spin field-effect switch}}.
\newblock \emph{\bibinfo{journal}{Nature Communications}}
  \textbf{\bibinfo{volume}{7}}, \bibinfo{pages}{13372} (\bibinfo{year}{2016}).

\bibitem{Dankert2017}
\bibinfo{author}{Dankert, A.} \& \bibinfo{author}{Dash, S.~P.}
\newblock \bibinfo{title}{{Electrical gate control of spin current in van der
  Waals heterostructures at room temperature}}.
\newblock \emph{\bibinfo{journal}{Nature Communications}}
  \textbf{\bibinfo{volume}{8}}, \bibinfo{pages}{16093} (\bibinfo{year}{2017}).

\bibitem{Ringer2018}
\bibinfo{author}{Ringer, S.} \emph{et~al.}
\newblock \bibinfo{title}{Spin field-effect transistor action via tunable
  polarization of the spin injection in a co/mgo/graphene contact}.
\newblock \emph{\bibinfo{journal}{Applied Physics Letters}}
  \textbf{\bibinfo{volume}{113}}, \bibinfo{pages}{132403}
  (\bibinfo{year}{2018}).

\bibitem{Xu2018}
\bibinfo{author}{Xu, J.} \emph{et~al.}
\newblock \bibinfo{title}{{Spin inversion in graphene spin valves by
  gate-tunable magnetic proximity effect at one-dimensional contacts}}.
\newblock \emph{\bibinfo{journal}{Nature Communications}}
  \textbf{\bibinfo{volume}{9}}, \bibinfo{pages}{2869} (\bibinfo{year}{2018}).

\bibitem{Guarochico-Moreira2022}
\bibinfo{author}{Guarochico-Moreira, V.~H.} \emph{et~al.}
\newblock \bibinfo{title}{{Tunable Spin Injection in High-Quality Graphene with
  One-Dimensional Contacts}}.
\newblock \emph{\bibinfo{journal}{Nano Letters}} \textbf{\bibinfo{volume}{22}},
  \bibinfo{pages}{935--941} (\bibinfo{year}{2022}).

\bibitem{Avsar2016}
\bibinfo{author}{Avsar, A.} \emph{et~al.}
\newblock \bibinfo{title}{{Electronic spin transport in dual-gated bilayer
  graphene}}.
\newblock \emph{\bibinfo{journal}{NPG Asia Materials}}
  \textbf{\bibinfo{volume}{8}}, \bibinfo{pages}{e274} (\bibinfo{year}{2016}).

\bibitem{Ingla-Aynes2015}
\bibinfo{author}{Ingla-Ayn\'es, J.}, \bibinfo{author}{Guimar\~aes, M. H.~D.},
  \bibinfo{author}{Meijerink, R.~J.}, \bibinfo{author}{Zomer, P.~J.} \&
  \bibinfo{author}{van Wees, B.~J.}
\newblock \bibinfo{title}{$24\ensuremath{-}\ensuremath{\mu}\mathrm{m}$ spin
  relaxation length in boron nitride encapsulated bilayer graphene}.
\newblock \emph{\bibinfo{journal}{Phys. Rev. B}} \textbf{\bibinfo{volume}{92}},
  \bibinfo{pages}{201410} (\bibinfo{year}{2015}).

\bibitem{Ingla-Aynes2021}
\bibinfo{author}{Ingla-Ayn\'es, J.}, \bibinfo{author}{Herling, F.},
  \bibinfo{author}{Fabian, J.}, \bibinfo{author}{Hueso, L.~E.} \&
  \bibinfo{author}{Casanova, F.}
\newblock \bibinfo{title}{Electrical control of valley-zeeman
  spin-orbit-coupling--induced spin precession at room temperature}.
\newblock \emph{\bibinfo{journal}{Phys. Rev. Lett.}}
  \textbf{\bibinfo{volume}{127}}, \bibinfo{pages}{047202}
  (\bibinfo{year}{2021}).

\bibitem{Takahashi2003}
\bibinfo{author}{Takahashi, S.} \& \bibinfo{author}{Maekawa, S.}
\newblock \bibinfo{title}{Spin injection and detection in magnetic
  nanostructures}.
\newblock \emph{\bibinfo{journal}{Phys. Rev. B}} \textbf{\bibinfo{volume}{67}},
  \bibinfo{pages}{052409} (\bibinfo{year}{2003}).

\bibitem{Maassen2012}
\bibinfo{author}{Maassen, T.}, \bibinfo{author}{Vera-Marun, I.~J.},
  \bibinfo{author}{Guimar\~aes, M. H.~D.} \& \bibinfo{author}{van Wees, B.~J.}
\newblock \bibinfo{title}{Contact-induced spin relaxation in hanle spin
  precession measurements}.
\newblock \emph{\bibinfo{journal}{Phys. Rev. B}} \textbf{\bibinfo{volume}{86}},
  \bibinfo{pages}{235408} (\bibinfo{year}{2012}).

\bibitem{Xu2018b}
\bibinfo{author}{Xu, J.}, \bibinfo{author}{Zhu, T.}, \bibinfo{author}{Luo,
  Y.~K.}, \bibinfo{author}{Lu, Y.-M.} \& \bibinfo{author}{Kawakami, R.~K.}
\newblock \bibinfo{title}{Strong and tunable spin-lifetime anisotropy in
  dual-gated bilayer graphene}.
\newblock \emph{\bibinfo{journal}{Phys. Rev. Lett.}}
  \textbf{\bibinfo{volume}{121}}, \bibinfo{pages}{127703}
  (\bibinfo{year}{2018}).

\bibitem{Tombros2007}
\bibinfo{author}{Tombros, N.}, \bibinfo{author}{Jozsa, C.},
  \bibinfo{author}{Popinciuc, M.}, \bibinfo{author}{Jonkman, H.~T.} \&
  \bibinfo{author}{van Wees, B.~J.}
\newblock \bibinfo{title}{{Electronic spin transport and spin precession in
  single graphene layers at room temperature}}.
\newblock \emph{\bibinfo{journal}{Nature}} \textbf{\bibinfo{volume}{448}},
  \bibinfo{pages}{571--574} (\bibinfo{year}{2007}).

\bibitem{Han2010}
\bibinfo{author}{Han, W.} \emph{et~al.}
\newblock \bibinfo{title}{Tunneling spin injection into single layer graphene}.
\newblock \emph{\bibinfo{journal}{Phys. Rev. Lett.}}
  \textbf{\bibinfo{volume}{105}}, \bibinfo{pages}{167202}
  (\bibinfo{year}{2010}).

\bibitem{Gurram2016a}
\bibinfo{author}{Gurram, M.} \emph{et~al.}
\newblock \bibinfo{title}{Spin transport in fully hexagonal boron nitride
  encapsulated graphene}.
\newblock \emph{\bibinfo{journal}{Phys. Rev. B}} \textbf{\bibinfo{volume}{93}},
  \bibinfo{pages}{115441} (\bibinfo{year}{2016}).

\bibitem{Leutenantsmeyer2018}
\bibinfo{author}{Leutenantsmeyer, J.~C.}, \bibinfo{author}{Ingla-Ayn\'es, J.},
  \bibinfo{author}{Fabian, J.} \& \bibinfo{author}{van Wees, B.~J.}
\newblock \bibinfo{title}{Observation of spin-valley-coupling-induced large
  spin-lifetime anisotropy in bilayer graphene}.
\newblock \emph{\bibinfo{journal}{Phys. Rev. Lett.}}
  \textbf{\bibinfo{volume}{121}}, \bibinfo{pages}{127702}
  (\bibinfo{year}{2018}).

\bibitem{Gorbachev2011}
\bibinfo{author}{Gorbachev, R.~V.} \emph{et~al.}
\newblock \bibinfo{title}{Hunting for monolayer boron nitride: Optical and
  raman signatures}.
\newblock \emph{\bibinfo{journal}{Small}} \textbf{\bibinfo{volume}{7}},
  \bibinfo{pages}{465--468} (\bibinfo{year}{2011}).

\bibitem{Johnson1988}
\bibinfo{author}{Johnson, M.} \& \bibinfo{author}{Silsbee, R.~H.}
\newblock \bibinfo{title}{Coupling of electronic charge and spin at a
  ferromagnetic-paramagnetic metal interface}.
\newblock \emph{\bibinfo{journal}{Phys. Rev. B}} \textbf{\bibinfo{volume}{37}},
  \bibinfo{pages}{5312--5325} (\bibinfo{year}{1988}).

\bibitem{Arrhenius1889a}
\bibinfo{author}{Arrhenius, S.}
\newblock \bibinfo{title}{{{\"{U}}ber die Dissociationsw{\"{a}}rme und den
  Einfluss der Temperatur auf den Dissociationsgrad der Elektrolyte}}.
\newblock \emph{\bibinfo{journal}{Zeitschrift f{\"{u}}r Physikalische Chemie}}
  \textbf{\bibinfo{volume}{4U}}, \bibinfo{pages}{96--116}
  (\bibinfo{year}{1889}).

\bibitem{Arrhenius1889}
\bibinfo{author}{Arrhenius, S.}
\newblock \bibinfo{title}{{{\"{U}}ber die Reaktionsgeschwindigkeit bei der
  Inversion von Rohrzucker durch S{\"{a}}uren}}.
\newblock \emph{\bibinfo{journal}{Zeitschrift f{\"{u}}r Physikalische Chemie}}
  \textbf{\bibinfo{volume}{4U}}, \bibinfo{pages}{226--248}
  (\bibinfo{year}{1889}).

\bibitem{Fabian2007}
\bibinfo{author}{Fabian, J.}, \bibinfo{author}{Matos-Abiague, A.},
  \bibinfo{author}{Ertler, C.}, \bibinfo{author}{Stano, P.} \&
  \bibinfo{author}{{\v{Z}}uti{\'{c}}, I.}
\newblock \bibinfo{title}{{Semiconductor spintronics}}.
\newblock \emph{\bibinfo{journal}{Acta Physica Slovaca. Reviews and Tutorials}}
  \textbf{\bibinfo{volume}{57}} (\bibinfo{year}{2007}).

\bibitem{Konschuh2012}
\bibinfo{author}{Konschuh, S.}, \bibinfo{author}{Gmitra, M.},
  \bibinfo{author}{Kochan, D.} \& \bibinfo{author}{Fabian, J.}
\newblock \bibinfo{title}{Theory of spin-orbit coupling in bilayer graphene}.
\newblock \emph{\bibinfo{journal}{Phys. Rev. B}} \textbf{\bibinfo{volume}{85}},
  \bibinfo{pages}{115423} (\bibinfo{year}{2012}).

\bibitem{Guimaraes_Controlling_2014}
\bibinfo{author}{Guimar\~aes, M. H.~D.} \emph{et~al.}
\newblock \bibinfo{title}{Controlling spin relaxation in hexagonal
  bn-encapsulated graphene with a transverse electric field}.
\newblock \emph{\bibinfo{journal}{Phys. Rev. Lett.}}
  \textbf{\bibinfo{volume}{113}}, \bibinfo{pages}{086602}
  (\bibinfo{year}{2014}).

\bibitem{Sierra2021}
\bibinfo{author}{Sierra, J.~F.}, \bibinfo{author}{Fabian, J.},
  \bibinfo{author}{Kawakami, R.~K.}, \bibinfo{author}{Roche, S.} \&
  \bibinfo{author}{Valenzuela, S.~O.}
\newblock \bibinfo{title}{{Van der Waals heterostructures for spintronics and
  opto-spintronics}}.
\newblock \emph{\bibinfo{journal}{Nature Nanotechnology}}
  \textbf{\bibinfo{volume}{16}}, \bibinfo{pages}{856--868}
  (\bibinfo{year}{2021}).

\bibitem{Yang2022c}
\bibinfo{author}{Yang, H.} \emph{et~al.}
\newblock \bibinfo{title}{{Two-dimensional materials prospects for non-volatile
  spintronic memories}}.
\newblock \emph{\bibinfo{journal}{Nature}} \textbf{\bibinfo{volume}{606}},
  \bibinfo{pages}{663--673} (\bibinfo{year}{2022}).

\bibitem{Ingla-Aynes2018a}
\bibinfo{author}{Ingla-Ayn{\'{e}}s, J.}, \bibinfo{author}{Kaverzin, A.~A.} \&
  \bibinfo{author}{van Wees, B.}
\newblock \bibinfo{title}{{Carrier Drift Control of Spin Currents in
  Graphene-Based Spin-Current Demultiplexers}}.
\newblock \emph{\bibinfo{journal}{Physical Review Applied}}
  \textbf{\bibinfo{volume}{10}}, \bibinfo{pages}{044073}
  (\bibinfo{year}{2018}).

\bibitem{Novoselov2004}
\bibinfo{author}{Novoselov, K.~S.} \emph{et~al.}
\newblock \bibinfo{title}{Electric field effect in atomically thin carbon
  films}.
\newblock \emph{\bibinfo{journal}{Science}} \textbf{\bibinfo{volume}{306}},
  \bibinfo{pages}{666--669} (\bibinfo{year}{2004}).

\end{thebibliography}

\section*{Acknowledgments}
Financial support by the Engineering and Physical Science Research Council (EPSRC) for a PhD studentship and Doctoral Prize Fellowship is gratefully acknowledged, as well as support from the Graphene Flagship Core3 project. Additionally, the 
financial support of Consejo Nacional de Ciencia y Tecnología (CONACyT), México for a PhD studentship is also gratefully acknowledged.

\section*{Author Contributions}
C.R.A. and N.N. carried out the research. V. H. G. provided assistance with the device fabrication. The work was supervised by I.V.G. and I.J.V. All authors contributed to the preparation of the manuscript.

\section*{Additional Information}
\textbf{Competing Interests:} The authors declare no competing interests.

\end{document}